\documentclass[conference]{IEEEtran}
\IEEEoverridecommandlockouts
\usepackage{cite}
\usepackage{amsmath,amssymb,amsfonts}
\usepackage{bbm}
\usepackage{algorithm,algorithmic}
\usepackage{graphicx}
\usepackage{textcomp}
\usepackage{xcolor}
\def\BibTeX{{\rm B\kern-.05em{\sc i\kern-.025em b}\kern-.08em
    T\kern-.1667em\lower.7ex\hbox{E}\kern-.125emX}}
\usepackage{float}
\usepackage{amsmath}
\usepackage{caption}
\newtheorem{theorem}{Theorem}
\newtheorem{lemma}{Lemma}
\newtheorem{definition}{Definition}

\newtheorem{pb}{Problem}

\graphicspath{{figures/}}
\allowdisplaybreaks
\begin{document}

\title{Optimizing Data Freshness in Time-Varying Wireless Networks with Imperfect Channel State}
\author{
	\IEEEauthorblockN{Haoyue~Tang\textsuperscript{1},~Jintao~Wang\textsuperscript{1,2},~Philippe~Ciblat\textsuperscript{3},~Jian~Song\textsuperscript{1,2}}
	\IEEEauthorblockA{
		\textsuperscript{1}Beijing National Research Center for Information Science and Technology (BNRist),\\
		Dept. of Electronic Engineering, Tsinghua University, Beijing 100084, China\\
		\textsuperscript{2}Research Institute of Tsinghua University in Shenzhen, Shenzhen, 518057\\
		\textsuperscript{3}Telecom ParisTech, Institut Polytechnique de Paris, Paris, France\\
		\{thy17@mails, wangjintao@,jsong@\}tsinghua.edu.cn, ciblat@telecom-paristech.fr}}

\maketitle

\begin{abstract}
	We consider a scenario where a base station (BS) attempts to collect fresh information from power constrained sensors over time-varying band-limited wireless channels. We characterize the data freshness through the recently proposed metric -- the \emph{Age of Information}. We consider a time-varying channel model with power adaptation. Unlike previous work, packet loss may happen due to imperfect channel estimation or decoding error. We propose an asymptotic optimal scheduling algorithm minimizing AoI performance and satisfying both bandwidth and power constraint in such networks. Numerical simulations show that the proposed policy outperforms the greedy one, and we observe that sensors with poor channels are scheduled at higher AoI in order to limit the packet loss.  
\end{abstract}


\section{Introduction}
 \let\thefootnote\relax\footnotetext{\noindent -----------------\\ This work was supported by the National Key R\&D Program of China (Grant No.2017YFE0112300), Shenzhen basic Research Project (No.JCYJ20170816152246879), the ERC (Grant No. 715111) and the Tsinghua University Tutor Research Fund.}
Nowadays, the proliferation of real-time applications like virtual reality and the Internet of things (IoT) networks imposes new requirement on data freshness. Moreover, the unprecedented growth of the number of sensors and transmitters in such systems creates challenge to scheduling and routing strategy design under a limited bandwidth. However, previous solutions that can achieve a high throughput and low delay performance may not satisfy a good data freshness requirement. Thus, to meet the aforementioned challenges, studying scheduling strategies that aim at optimizing data freshness is an important issue.

To measure data freshness from the perspective of receiver, the metric called ``Age of Information'', namely the difference between now and the freshest information at the receiver is generated, has been proposed \cite{yates12infocom}. Since then, scheduling strategies that minimize the AoI performance under limited communication resources such as bandwidth and power have received lots of attention, see e.g.\cite{igor_18_ton,jiang_isit_2018,lu_age_2018,hsu_scheduling_2018,qing18tit,bo_18_GC}. In \cite{lu_age_2018}, time-varying transmission rate is allowed and scheduling policy based on Lyapunov optimization is proposed. In \cite{collins1999transmission,uysal_02_ton,wang_19_tcomm}, cross-layer control strategy via time-varying transmission power has been proposed to optimize transmission throughput and delay performance with power constraint. In our previous work \cite{haoyue}, cross-layer control strategy is also applied to minimize AoI but when each transmitter is restricted to an average power constraint. Moreover, it has been done under a restrictive assumption that perfect Channel State Information at the Transmitter (CSIT) is available and no transmission error occurs.

In this paper, we consider the problem of multi-user scheduling to minimize AoI in a bandwidth limited error-prone time-varying networks. Unlike the aforementioned papers that consider packet transmission to be perfect \cite{lu_age_2018,haoyue,wang_19_tcomm}, we assume packet-loss may happen due to the imperfect estimated channel states or decoding error, which is a common phenomenon in wireless communications. We first decouple the multi-user scheduling problem into a single-user cost minimization problem and formulate it into a constrained Markov decision process. By taking the packet-loss into consideration, we adopt a linear programming (LP) approach to figure out the optimum transmission strategy for the decoupled single user. Finally, a multi-user scheduling strategy satisfying the bandwidth constraint is proposed and we prove that its performance is optimal asymptotically (with respect to the number of sensors).

The rest of the paper is organized as follows: We introduce the system model and state the scheduling problem in Section II. In Section III, we first decouple the multi-user scheduling problem into single-user constrained Markov decision process and solve it through LP. A practical scheduling algorithm is then proposed to satisfy the bandwidth constraint. Section IV provides the simulation results and Section V draws the conclusion. 

\emph{Notations:} Matrices are written in bold upper letters. The cardinality of a set $\mathcal{S}$ is denoted as $|\mathcal{S}|$. Let the upper and lower letters denote random variables and their realizations, respectively. The expectation of random variable $X$ is denoted by $\mathbb{E}[X]$ and the probability of event $A$ is denoted by $\text{Pr}(A)$. 
\section{System Model}

\subsection{Network Model}

\begin{figure}[h]
	\centering
	\includegraphics[width=.25\textwidth]{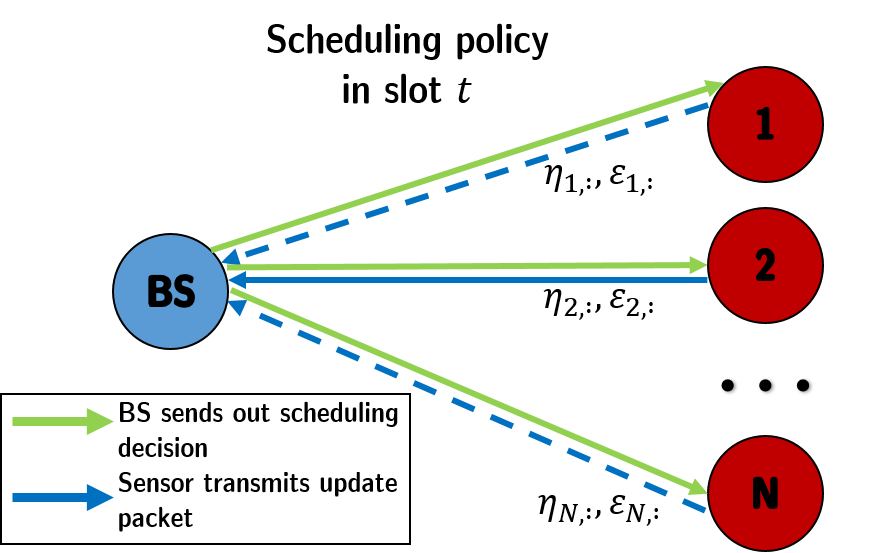}
	\caption{A bandwidth limited multi-user network. }
	\label{fig:network}
\end{figure}

In this work, we consider scheduling strategy for a base station (BS) collecting fresh data from $N$ power constrained sensors over bandwidth limited time-varying channels, as is depicted in Fig.~\ref{fig:network}. Let us consider a discrete time scenario and let $t\in\mathbb{Z}^+$ denote the index of slot. Due to bandwidth and interference constraint, at the beginning of each slot, the BS can schedule no more than $M$ sensors to upload update packets. Let $u_n(t)\in\{0, 1\}$ denote scheduling decisions. If $u_n(t)=1$, then sensor $n$ is scheduled and sends information he observes at the beginning of slot $t$, the packet will be received at the end of slot $t$ if the transmission succeeds. 

\begin{figure}[h]
	\centering
	\includegraphics[width=.35\textwidth]{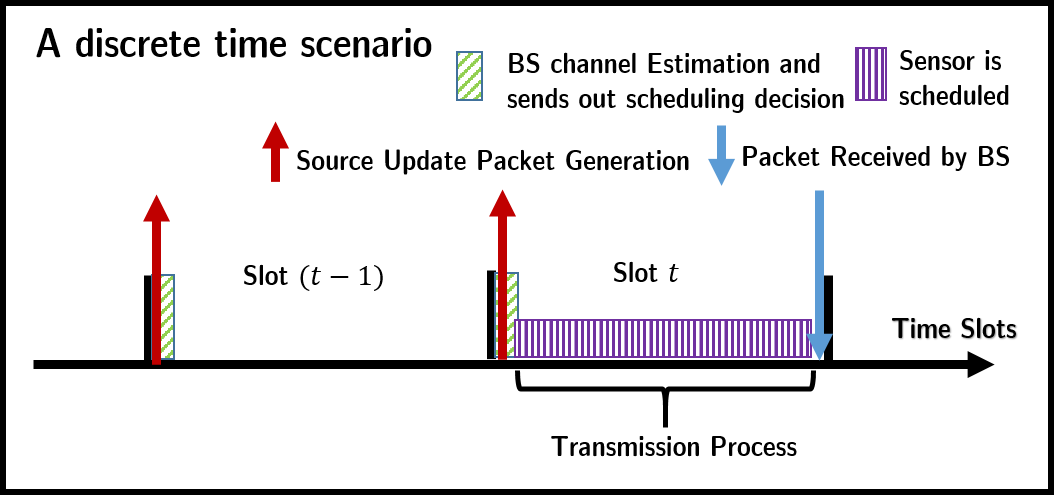}
	\caption{Illustration transmission decision for sensor $2$}
	\label{fig:timeslot}
\end{figure}

Let $q_n(t)\in\{1, \cdots, Q\}$ be the estimated channel state information (CSI) between sensor $n$ and the BS at slot $t$. At the beginning of each slot, each sensor wakes up and sends a tiny pilot sequence for the BS to have an estimate $q_n(t)$ about the channel state, as is depicted in Fig.~\ref{fig:timeslot}. The estimated channel state is then sent back to each sensor in combination with the scheduling decisions. To combat wireless fading, sensor $n$ will consume an amount of $\omega_{q_n(t)}$ power to transmit updates if chosen to be scheduled in slot $t$. Larger $q$ stands for more noisy channels that goes through stronger fading, thus $\omega_1<\cdots<\omega_Q$ is a sequence in ascending order. For scheduling decisions $[u_n(1), \cdots, u_n(T)]$ of sensor $n$ over a consecutive of $T$ slots, the average power consumed must satisfy its own power constraint, i.e.,
\begin{equation}
	\frac{1}{T}\sum_{t=1}^Tu_n(t)\omega_{q_n(t)}\leq\mathcal{E}_n. 
\end{equation}

We assume if the estimated channel $q_n(t)=q$, there is a probability of $\varepsilon_{n, q}$ that the transmitted packet cannot be received successfully at the BS. We assume for each sensor, channel state varies independently across the slot, thus the estimated value $q$ appears independently in each slot with probability $\eta_{n, q}$, the sum of which satisfies:
\begin{equation}
	\sum_{q=1}^Q\eta_{n, q}=1, \forall n. \label{eq:probabilityeta}
\end{equation}

\subsection{Age of Information}
In this work, we use \emph{Age of Information} \cite{yates12infocom} to measure data freshness from the perspective of BS. Denote $x_n(t)$ be the AoI of sensor $n$ at the beginning of slot $t$. By definition, AoI measures the difference between now and the time freshest information at the BS has been created. If $u_n(t)=1$ and the estimated channel state is $q_n(t)$, the transmission from sensor $n$ to the BS succeeds with probability $1-\varepsilon_{n, q_n(t)}$, then $x_n(t+1)=1$; otherwise, if the transmission fails with probability $\varepsilon_{n, q_n(t)}$, or sensor $n$ is not scheduled in slot $t$, then the AoI of sensor $n$ will be one slot older at the beginning of slot $t+1$, i.e., $x_n(t+1)=x_n(t)+1$. Based on the above analysis, the AoI evolution relationship is provided as follows:
\begin{align}
	\text{Pr}(&x_n(t+1)|x_n(t), u_n(t))\nonumber\\=&\begin{cases}
	1-\varepsilon_{n, q_n(t)}, &x_n(t+1)=1, u_n(t)=1;\\
	\varepsilon_{n, q_n(t)}, &x_n(t+1)=x_n(t)+1, u_n(t)=1;\\
	1, &x_n(t+1)=x_n(t)+1, u_n(t)=0;\\
	0, &\text{otherwise. }
	\end{cases}
	\label{eq:probabilitytransfer}
\end{align}
\subsection{Problem Formulation}
Suppose the channel statistics $\eta_{n, q}$, the estimation error $\varepsilon_{n, q}$ and the average power constraint $\mathcal{E}_n$ of each user $n$ is known at the BS. The goal is to design a scheduling algorithm for the BS such that the average AoI at the beginning of each slot can be minimized under both bandwidth and power constraints. The scheduling decisions should be made based on channel and power constraint statistics, as well as the current AoI $x_n(t)$ and the estimate $q_n(t)$, while no prediction about the future AoI or channel states can be used. Denote the set of such scheduling decisions as $\Pi_{\text{NA}}$, the mathematic problem formulation is provided as follows:
\begin{pb}[Primal Scheduling Problem]
\begin{subequations}
	\begin{equation}
		\text{AoI}^{\text{opt}}=\min_{\pi\in\Pi_{\text{NA}}}\lim_{T\rightarrow\infty}\mathbb{E}_{\pi}\left[\frac{1}{NT}\sum_{t=1}^T\sum_{n=1}^Nx_n(t)\right],
	\end{equation}

	\begin{align}
		\text{s.t. }&\sum_{n=1}^Nu_n(t)\leq M,\forall t,\label{eq:bandwidth}\\
		&\mathbb{E}_{\pi}\left[\frac{1}{T}\sum_{t=1}^Tu_n(t)\omega_{q_n(t)}\right]\leq\mathcal{E}_n, \forall n.\label{eq:powerconstraint} 
	\end{align}
\end{subequations}
\end{pb}
\section{Problem Resolution}
\subsection{Decoupled Single Sensor Problem Reformulation}
Notice that $u_n(t)$ can take only 0 and 1, the primal scheduling problem is an NP-hard integer programming due to constraint Eq.~\eqref{eq:bandwidth}. To overcome this challenge, we first relax the hard constraint in every slot into a time average bandwidth constraint. After relaxation, the number of sensors scheduled in each slot can be larger than $M$, but the time average sensors scheduled per slot is still smaller than $M$. 
\begin{equation}
	\mathbb{E}_{\pi}\left[\frac{1}{T}\sum_{t=1}^T\sum_{n=1}^Nu_n(t)\right]\leq M. \label{eq:relaxedbandwidth}
\end{equation}

To obtain the decoupled scheduling problem of each sensor, the relaxed bandwidth constraint is taken into consideration through Lagrange reformulation. Let $C\geq 0$ be the Lagrange multiplier and the Lagrange function can be written out as follows:
\begin{equation}
	\mathcal{L}(\pi, C)=\mathbb{E}_\pi\left[\frac{1}{NT}\sum_{t=1}^T\sum_{n=1}^N(x_n(t)+Cu_n(t))-\frac{CM}{N}\right].\label{eq:langrange}
\end{equation}

Denote $\pi_C$ be the optimum scheduling policy that minimizes the above Lagrange function for given multiplier $C$. The optimum policy that minimizes the average AoI with relaxed bandwidth Eq.~\eqref{eq:relaxedbandwidth} is a mixture of no more than two $\pi_C$ correspond to different multipliers. The selection of multiplier $C$ will be explained in Sec.~III-B. Let us first compute the optimum policy $\pi_C$ for a given $C$. Notice that the minimizing Eq.~\eqref{eq:langrange} has no bandwidth constraint, thus it can be decoupled into $N$ subproblems as follows:
\begin{pb}[Decoupled Cost Minimization with Power Constraint]
	\begin{subequations}
		\begin{equation}
			\pi_n^*=\arg\min_{\pi\in\Pi_{\text{NA}}}\mathbb{E}_\pi\left[\frac{1}{T}\sum_{t=1}^T(x_n(t)+Cu_n(t))\right],
		\end{equation}
		\begin{equation}
			\text{s.t. }\mathbb{E}_\pi\left[\frac{1}{T}\sum_{t=1}^Tu_n(t)\omega_{q_n(t)}\right]\leq\mathcal{E}_n. 
		\end{equation}
	\end{subequations}
\end{pb}

Then each of the problem can be solved separately and the subscript $n$ is omitted henceforth. 
\subsection{Constrained Markov Decision Process}
For fixed Lagrange multiplier $C$, each of the optimization problem can be formulated into a constrained Markov decision process (CMDP). The state space of the CMDP contains the AoI $x(t)$ at the beginning of slot $t$ and the estimate $q(t)$. The action $a(t)\in\{0, 1\}$ represents the scheduling decision for the decoupled problem. The probability transfer function for the MDP can be obtained through Eq.~\eqref{eq:probabilitytransfer}. The cost in each slot contains both AoI growth and the extra cost of schedule a user, i.e., $x(t)+Ca(t)$. The goal of the CMDP is to obtain a small time-average cost under power constraint Eq.~(7b). 

According to \cite[Proposition 4.6.5]{Bertsekas_DP}, the optimal policy for such MDP without power constraint is a stationary one. To analyze the optimal structure of power constraint, let us provide the formal definition of a stationary randomized policy.

\begin{definition}
	Let $\Pi_{\text{SR}}$ be the set of stationary randomized policies. For state $(x(t)=x, q(t)=q)$, a stationary randomized policy chooses action $a(t)=1$ with probability $p_{x, q}$ regardless of $t$. 
\end{definition}

\begin{theorem}
	The optimum strategy to the decoupled single sensor cost minimization problem has a threshold structure. For each estimated channel state $q$, if the probability to schedule the sensor when the age is $x$ is non-zero, then for states $\text{AoI}>x$, it is always optimum to schedule the sensor when the estimated channel state is $q$. 
\end{theorem}
\begin{IEEEproof}The proof is similar to Lemma 1 in \cite{haoyue} and is hence omitted here.
\end{IEEEproof}
\subsection{Optimum Single Sensor Scheduling}
Let us now consider a stationary randomized policy, where sensor with AoI $x(t)=x$ and the estimate $q(t)=q$ is scheduled with probability $p_{x, q}$. Our job then is to figure out a set of optimum parameters $\{p_{x, q}^*\}$, such that for fixed $C$ the overall cost can be minimized. According to the threshold structure obtained from Theorem 1, the exists a large $X$, such that for any $x>X$, the optimum policy is to schedule the sensor regardless of the estimated channel state $q$. 

Let $\mu_x$ be the steady state distribution that AoI of the decoupled sensor equals $x$ when a stationary randomized policy $\{p_{x, q}\}$ is employed. Let $\alpha_x$ and $\beta_x$ be the transfer probability that the AoI evolves from $x$ to $x+1$ and from $x$ to $1$ separately. If $x>X$, the optimum decision is to schedule the sensor, thus AoI increases if the transmission fails, or decreases to 1 if the transmission succeeds. Thus the forward and backward transfer probability can be written out as follows:
\begin{subequations}
\begin{align}
\alpha_x&=\begin{cases}
\sum_{q=1}^Q\eta_q\varepsilon_q, &x\geq X;\\
\sum_{q=1}^Q\eta_q(p_{x, q}\varepsilon_q+1-p_{x, q}), &x<X,
\end{cases}\\
\beta_x&=\begin{cases}
\sum_{q=1}^Q\eta_q(1-\varepsilon_q), &x\geq X;\\
\sum_{q=1}^Q\eta_qp_{x, q}(1-\varepsilon_q), &x<X.
\end{cases}
\end{align}
\end{subequations}
\begin{lemma}
	The probability transfer matrix $\mathbf{U}$ between AoI states equal $1, \cdots, X$ can be written out as follows:
	\begin{equation}
		\mathbf{U}=\left[\begin{matrix}
		\beta_1, \cdots, \beta_{X-1}, &1\\
		\text{diag}(\alpha_1, \cdots, \alpha_{X-1}), &\mathbf{0}_{X-1}
		\end{matrix}\right].\label{eq:transferU}
	\end{equation}
	Then, the steady state distribution $\boldsymbol{\mu}=[\mu_1, \cdots, \mu_{X}]^T$ can be computed by solving the following linear equations:
	\begin{equation}
		\left[\begin{matrix}
		\mathbf{U}-\mathbf{I}_{X}\\
		\mathbf{1}_{X-1}^T, (\sum_{q=1}^Q\eta_q\varepsilon_q)^{-1}
		\end{matrix}\right]\boldsymbol{\mu}=\left[\begin{matrix}
		\mathbf{0}_{X}\\1
		\end{matrix}\right].\label{eq:Xmaxsum}
	\end{equation}
\end{lemma}

\begin{IEEEproof}
	Let us first analyze state $x>X$. Since the optimum decision is to schedule the sensor, we have:
	\begin{equation}
		\mu_{x}=\alpha_x\mu_{x-1}=(\sum_{q=1}^Q\eta_q\varepsilon_q)^{x-X}\mu_{X}. 
		\label{eq:mu>}
	\end{equation}
	For simplicity, let us denote $\gamma=\sum_{q=1}^Q\eta_q\varepsilon_q$. Then, for $x>X$, the backward probability $\beta_x$ can be simplified:
	\begin{equation}
		\beta_x=\sum_{q=1}^Q\eta_q(1-\varepsilon_q)\overset{(a)}{=}1-\sum_{q=1}^Q\eta_q\varepsilon_q=1-\gamma,
		\label{eq:beta>}
	\end{equation} 
	where equality (a) is obtained because of Eq.~\eqref{eq:probabilityeta}.
	
	Notice that the AoI will go to state $x=1$ if and only if the transmission succeeds, thus
	\begin{align}
		\mu_1&=\sum_{x=1}^{X-1}\mu_x\beta_x+\sum_{x=X}^\infty\mu_x\beta_x\nonumber\\
		&\overset{(a)}{=}\sum_{x=1}^{X-1}\mu_x\beta_x+\sum_{x=X}^\infty\mu_{X}\gamma^{x-X}\beta_x\nonumber\\
		&\overset{(b)}{=}\sum_{x=1}^{X-1}\mu_x\beta_x+\mu_{X}\frac{1}{1-\gamma}\beta_x\nonumber\\
		&\overset{(c)}{=}\sum_{x=1}^{X-1}\mu_x\beta_x+\mu_{X},\label{eq:transferU1}
	\end{align}
	where equality (a) is obtained due to relationship Eq.~\eqref{eq:mu>}, equality (b) is due to the fact that $\sum_{x=X}^\infty\gamma^{x-X}=\frac{1}{1-\gamma}$ and equality (c) can be obtained by Eq.~\eqref{eq:beta>}. According to Eq.~\eqref{eq:transferU1}, items on the first line of the transfer matrix $\mathbf{U}$, i.e., Eq.~\eqref{eq:transferU} can be obtained. The remaining items of $\mathbf{U}$ can be obtained easily through $\mu_{x+1}=\alpha_x\mu_x$. 
	
	Notice that the sum of all the probability distribution satisfies
	\begin{equation}
		\sum_{x=1}^\infty \mu_x=1. \label{eq:musum}
	\end{equation}
	
	By substituting Eq.~\eqref{eq:mu>} into the above equation, we will have
	\begin{align}
		\sum_{x=1}^\infty \mu_x&=\sum_{x=1}^{X-1}\mu_x+\sum_{x=0}^\infty(\sum_{q=1}^Q\eta_q\varepsilon_q)^x\mu_X\nonumber\\
		&=\sum_{x=1}^{X-1}\mu_x+\frac{1}{1-\gamma}\mu_{X}=1,
	\end{align}
	which leads to the steady state distribution Eq.~\eqref{eq:Xmaxsum}.
\end{IEEEproof}	
	Denote $y_{x, q}=\mu_x\eta_qp_{x, q}$ be the probability that the sensor in state $x(t)=x$, the estimated channel state $q(t)=q$ and the sensor is scheduled for transmission. The following theorem will enable us to transfer the decoupled single user CMDP, i.e., Problem 2, into a Linear Programming (LP):
	\begin{theorem}
		The decoupled cost minimization problem is equivalent to the following LP problem:
		\begin{subequations}
			\begin{align}
				\{\mu_x^*, y_{x, q}^*\}&=
				\arg\min\big(\sum_{x=1}^{X-1}x\mu_x+\frac{1}{1-\gamma}X\mu_X+\label{eq:LPobj}\\
				&\frac{\gamma}{(1-\gamma)^2}\mu_X+C\sum_{x=1}^{X-1}\sum_{q=1}^Qy_{x,q}+\frac{C}{1-\gamma}\mu_{X}\big)\nonumber
			\end{align}
			\begin{align}
				\text{s.t. }&\sum_{x=1}^{X-1}\mu_x+\frac{1}{1-\gamma}\mu_{X}=1,\label{eq:LPC1}\\
				&\mu_1=\sum_{x=1}^{X-1}\sum_{q=1}^Qy_{x, q}(1-\varepsilon_q)+\mu_{X},\label{eq:LPC2}\\
				&\mu_x=\mu_{x-1}-\sum_{q=1}^Qy_{x-1, q}(1-\varepsilon_q),\forall 1<x\leq X,\label{eq:LPC3}\\
				&\sum_{q=1}^Q\sum_{x=1}^{X-1}y_{x, q}\omega_q+\frac{1}{1-\gamma}\mu_{X}\sum_{q=1}^Q\eta_q\omega_q\leq\mathcal{E},\label{eq:LPC4}\\
				&0\leq\mu_x\leq 1,0\leq y_{x, q}\leq\mu_x\eta_q, \forall x, q.\label{eq:LPC5}
			\end{align}
		\end{subequations}
	\end{theorem}
	\begin{IEEEproof}
		First let us compute the overall cost containing both AoI and scheduling cost. The total AoI can be computed by:
		\begin{align}
			\sum_{x=1}^\infty x\mu_x
			\overset{(a)}{=}&\sum_{x=1}^{X-1} x\mu_x+\sum_{x=0}^\infty (x+X)\gamma^x\mu_{X}\nonumber\\
			\overset{(b)}{=}&\sum_{x=1}^{X-1}x\mu_x+\frac{1}{1-\gamma}X\mu_{X}+\frac{\gamma}{(1-\gamma)^2}\mu_{X},
			\label{eq:AoIcost}
		\end{align}
		where equality (a) can be obtained by substituting Eq.~\eqref{eq:mu>} into the equation and equality (b) is obtained because $\sum_{x=0}^\infty\gamma^x=\frac{1}{1-\gamma}$ and $\sum_{x=0}^\infty x\gamma^x=\frac{\gamma}{(1-\gamma)^2}$. Next, let us compute the proportion of time slots spent on scheduling:
		\begin{align}
			\sum_{x=1}^\infty\sum_{q=1}^Qy_{x, q}
			\overset{(a)}{=}&\sum_{x=1}^{X-1}\sum_{q=1}^Qy_{x, q}+\sum_{x=X}^\infty\mu_x\nonumber\\
			\overset{(b)}{=}&\sum_{x=1}^{X-1}\sum_{q=1}^Qy_{x, q}+\sum_{x=0}^\infty\mu_{X}\gamma^x\nonumber\\
			=&\sum_{x=1}^{X-1}\sum_{q=1}^Qy_{x, q}+\frac{1}{1-\gamma}\mu_{X},
			\label{eq:schedulecost}
		\end{align}
		where equality (a) is obtained by the threshold structure and equality (b) is due to Eq.~\eqref{eq:mu>}. Summing up both Eq.~\eqref{eq:AoIcost} and Eq.~\eqref{eq:schedulecost} will yield the objective function Eq.~\eqref{eq:LPobj}. Constraint Eq.~\eqref{eq:LPC1}-\eqref{eq:LPC3} can be obtained through Eq.~\eqref{eq:Xmaxsum}. Next, we proceeds to compute the average power consumed by employing policy $\{\mu_x, y_{x, q}\}$:
		\begin{align}
			&\sum_{x=1}^{\infty}\sum_{q=1}^Qy_{x, q}\omega_q\nonumber\\
			=&\sum_{x=1}^{X-1}\sum_{q=1}^Qy_{x, q}\omega_q+\sum_{x=X}^\infty\sum_{q=1}^Q\mu_x\eta_q\omega_q\nonumber\\
			=&\sum_{x=1}^{X-1}\sum_{q=1}^Qy_{x, q}\omega_q+\sum_{q=1}^Q\eta_q\omega_q\mu_{X}\sum_{x=0}^\infty\gamma^x\nonumber\\
			=&\sum_{x=1}^{X-1}\sum_{q=1}^Qy_{x, q}\omega_q+\frac{1}{1-\gamma}\sum_{q=1}^Q\eta_q\omega_q\mu_X,
		\end{align}
		this yields constraint Eq.~\eqref{eq:LPC4}. Constraint Eq.~\eqref{eq:LPC5} is due to definition that $y_{x, q}=\mu_x\eta_qp_{x, q}$ and $p_{x, q}\leq 1$. 
	\end{IEEEproof}
	
	From the above theorem, for fixed $C$, the optimum scheduling decision for a decoupled single sensor can be obtained. Let $\{\mu_x(C), y_{x, q}(C)\}$ be the optimizer to the LP and denote $b(C)$ be the proportion of slots spent on scheduling the sensor, which can be computed as follows: 
	\begin{align}
		b(C)=\sum_{x=1}^{\infty}\sum_{q=1}^Qy_{x, q}(C)=\sum_{x=1}^{X-1}\sum_{q=1}^Qy_{x, q}(C)+\frac{1}{1-\gamma}\mu_X(C).
		\label{eq:bandwidthLP}
	\end{align}
	
	After compute the optimum scheduling decision when Lagrange multiplier equals $C$ separately for each decoupled sensor, denote $\{\mu_x^n(C), y_{x, q}^n(C)\}$ be the AoI and scheduling probability distribution of sensor $n$ under the optimum policy. Next, we will study how to obtain the optimum multipliers $C$.
	\subsection{Determination of Lagrange Multiplier $C$}
	The optimum policy $\pi_R^*$ that minimizes the average AoI under relaxed bandwidth Eq.~\eqref{eq:relaxedbandwidth} is a mixture of no more than two optimum policies that solves Problem 2 under different multipliers. Denote $\mu_x^{n, *}$ be the probability that the AoI of sensor $n$ equals $x$ under the optimum policy $\pi_R^*$ and let $y_{x, q}^{n, *}$ be the probability that the estimated channel state is $q$ and the sensor is scheduled. In this part, we explain how to obtain the two multipliers such that the relaxed constraint Eq.~\eqref{eq:relaxedbandwidth} can be satisfied. 
	
	We take a dual method to search for the multiplier. Suppose after the $k$-th iteration, the Lagrange multiplier is $C^{(k)}$. We compute the consumed bandwidth $b_n(C^{(k)})$ for each sensor $n$ by using Eq.~\eqref{eq:bandwidthLP}. The subgradient can be computed by:
	\begin{equation}
		d^{(k)}=\sum_{n=1}^Nb_n(C^{(k)})-M. 
	\end{equation}
	
	Let $s^{(k)}$ be a sequence of descending stepsize. if $d^{(k-1)}\cdot d^{(k)}<0$, we then smaller our descending stepsize by:
	\[
		s^{(k)}=\delta s^{(k-1)}, \delta\in(0, 1).
	\]
	
	The Lagrange multiplier used in the $(k+1)$-th iteration can then by computed by:
	\begin{equation}
		C^{(k+1)}=C^{(k)}+s^{(k)}d^{(k)}. 
	\end{equation}
	
	We start with $C^{(1)}=0$ and if $\sum_{n=1}^Nb_n(C^{(1)})\leq M$, indicating the relaxed bandwidth can satisfy all the power-constrained sensors. In this case, the optimum distribution can be obtained directly from
	\[\{\mu_{x}^{n, *}, y_{x, q}^{n, *}\}=\{\mu_x^n(C), y_{x, q}^n(C)\}.\]
	
	Otherwise, we first obtain a Lagrange multiplier sequence $C^{(k)}$ by following the above dual method. The iteration terminates until the step size is below a chosen threshold $s^{(k)}<\epsilon$. Next, we choose two items from the obtained sequence:
	\begin{subequations}
	\begin{align}
		C_u=\min_{k}\{C^{(k)}|\sum_{n=1}^Nb_n(C^{(k)})\geq M\},\\
		C_l=\max_{k}\{C^{(k)}|\sum_{n=1}^Nb_n(C^{(k)})<M\}.
	\end{align}
	\end{subequations}
	
	According to our previous analysis, the optimum scheduling policy of scheduling with relaxed bandwidth constraint $\pi_R$ is a mixture between policies obtained by the two Lagrange multipliers. Thus, the distribution $\{\mu_x^{n, *}, y_{x, q}^{n, *}\}$ can be obtained by computing a weighed average of $\{\mu_x^{n}(C_u), y_{x, q}^{n}(C_u)\}$ and $\{\mu_x^{n}(C_l), y_{x, q}^{n}(C_l)\}$ as follows:
	\begin{equation}
		\{\mu_x^{n, *}, y_{x, q}^{n, *}\}\!=\!\lambda\{\mu_x^n(C_u), y_{x, q}^n(C_u)\}+(1\!-\!\lambda)\{\mu_x^n(C_l
		), y_{x, q}^n(C_l)\},
	\end{equation}
	where the coefficient $\lambda$ can be computed by:
	\begin{equation}
		\lambda=\frac{M-\sum_{n=1}^Nb_n(C_l)}{\sum_{n=1}^Nb_n(C_u)-\sum_{n=1}^Nb_n(C_l)}.
	\end{equation}
	\subsection{Proposed Scheduling Algorithm}
	According to the threshold structure in Theorem 1, under the relaxed bandwidth constraint Eq.~\eqref{eq:relaxedbandwidth}, we can first compute the optimum probability $p_{x, q}^{n, *}$ to schedule sensor $n$ with AoI equals $x$ and in estimated channel state $q$. Theorem 1 implies for AoI larger than the threshold $X$, the optimum choice is to schedule the sensor to send updates. Thus, $p_{x, q}^{n, *}$ can be computed by:
	\begin{equation}
		p_{x, q}^{n, *}=\begin{cases}
		\frac{y_{x, q}^{n, *}}{\mu_x^{n, *}\eta_{n, q}}, &x\leq X\\
		1,&x>X.
		\end{cases}
	\end{equation}
	
	If there is only a single sensor in the network, i.e., $N=1$, then bandwidth constraint is not a problem. Scheduling the single sensor with probability $p_{x, q}^{1, *}$ when his AoI equals $x$ and in estimate $q$ is the optimum strategy that can minimize AoI under power constraint. When multiple power constrained users share a bandwidth limited network, we propose a truncated scheduling policy $\hat{\pi}$ based on the optimum scheduling probability of each sensor. At the beginning of each slot $t$, we first observe the AoI $x_n(t)$ and the estimated channel state $q_n(t)$ of user $n$. Next, put the index of sensor $n$ into the set $\mathcal{I}(t)$ with probability $p_{x_n(t), q_n(t)}^{n, *}$. If $|\mathcal{I}(t)|\leq M$, schedule all the sensors in set $\mathcal{I}(t)$ at the beginning of slot $t$; otherwise, choose a subset of $M$ sensors randomly from $\mathcal{I}(t)$ to carry out schedule decisions. 

\begin{theorem}
	The proposed scheduling policy $\hat{\pi}$ is asymptotic optimum. Denote $J(\pi)$ and the expected average AoI performance for following policy $\pi$ and let $\pi^*$ be the optimum scheduling policy to the primal scheduling problem with hard bandwidth constraint, keep $M/N$ as a constant, 
	\begin{equation}
		\lim_{N\rightarrow\infty}J(\hat{\pi})-J(\pi^*)\rightarrow0.
	\end{equation}
\end{theorem}
\begin{IEEEproof}
	See the Appendix for detailed proof. 
\end{IEEEproof}

\section{Simulations}
In this part we first study scheduling decision and the AoI performance for a power constrained single sensor and study how channel state imperfection affect our scheduling decisions. Next, we study the performance of our proposed algorithm in a multi-user network. 
\subsection{Single Sensor Scheduling Strategy}
\begin{figure}[h]
	\centering
	\includegraphics[width=.35\textwidth]{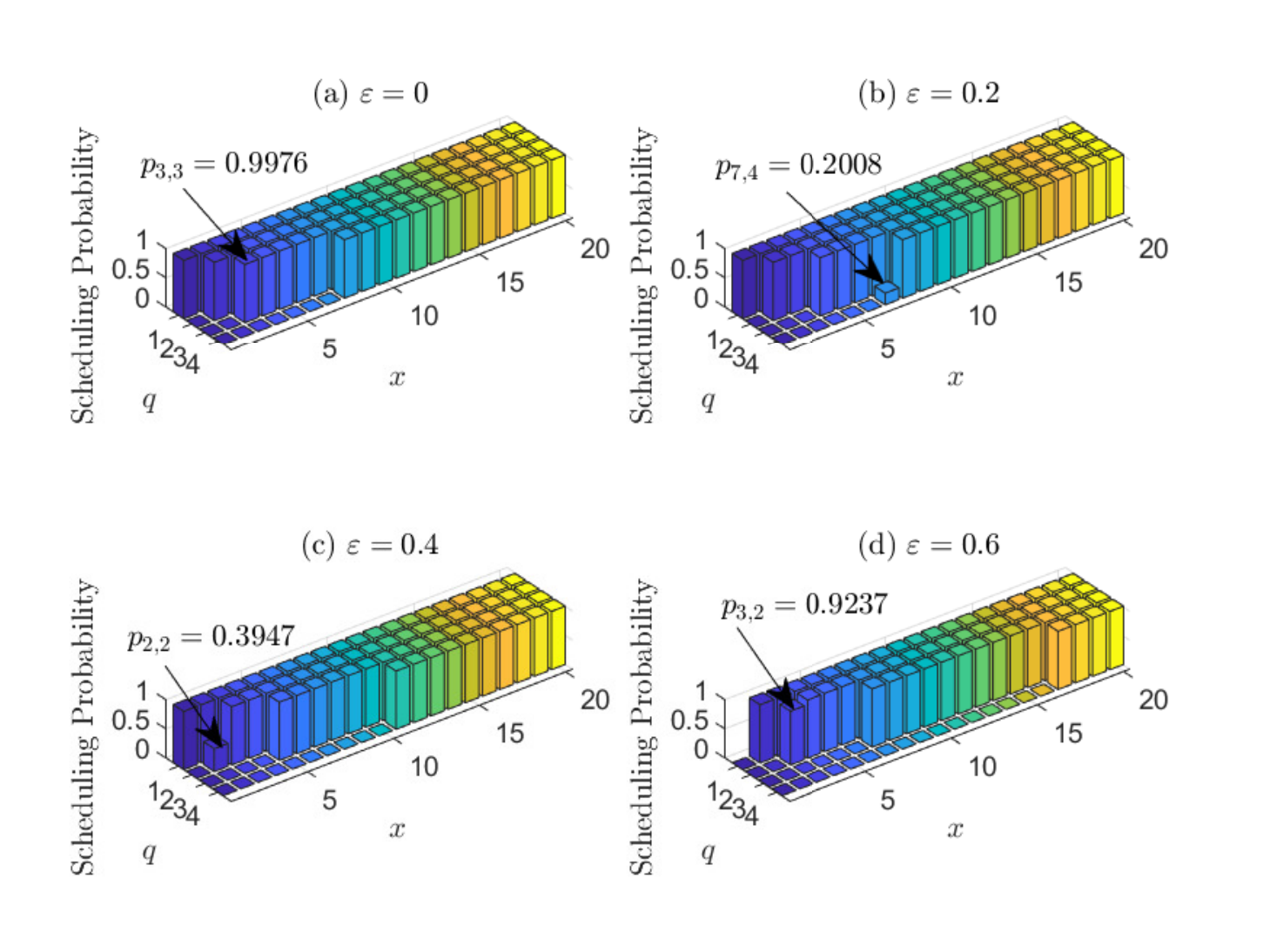}
	\caption{Optimum scheduling strategy for a single power constrained user with different probability of packet-loss.}
	\label{fig:strategy}
\end{figure}

We consider a $Q=4$ states channel, and the transmission power for each estimated $q$ is set to be $\omega_q=2^q$. We assume each estimate appears with identical probability $\eta_{n,q}=0.25, \forall n, q$ and the user has an average power constraint of $\mathcal{E}=0.5\sum_{q=1}^Q\eta_q\omega_q$. The probability of packet-loss is the same in each estimated channel state and equals $\varepsilon$. Fig.~\ref{fig:strategy} plots the scheduling strategy under different packet-loss and Table~\ref{tab:strategy} displays their average AoI performance. The strategy verifies the threshold structure presented in Theorem 1. Moreover, scheduling thresholds of all the states increase when the probability of packet-loss increases. Scheduling in higher packet-loss scenario will inevitably cause high AoI. 
\begin{table}[h]
	\small
	\centering
\caption{Average AoI performance}	
\begin{tabular}{|c|c|c|c|c|}
\hline
	$\varepsilon$&0&0.2&.4&0.6\\
	\hline
	Average AoI&1.8518&2.2648&2.9795&4.3508\\
	\hline
\end{tabular}
\label{tab:strategy}
\end{table}

\subsection{Multiple Sensor Average AoI Performance}
\begin{figure}[h]
	\centering
	\includegraphics[width=.4\textwidth]{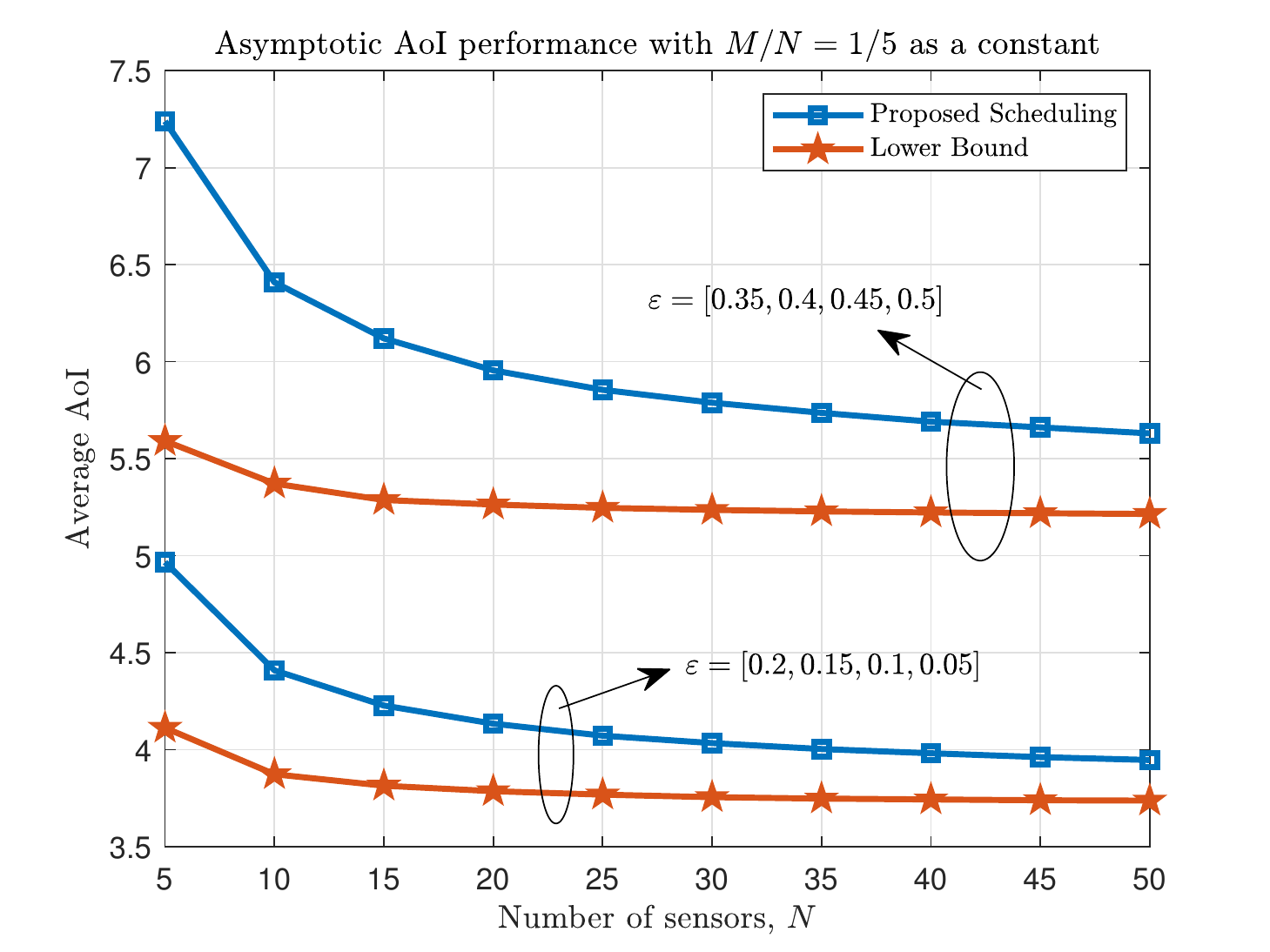}
	\caption{Asymptotic AoI performance under with $M/N=1/5$ be a constant under different channel states.}
	\label{fig:asymptotic}
\end{figure}

Fig.~\ref{fig:asymptotic} studies the asymptotic AoI performance with $M/N$ fixed as a constant. We consider a network where all the sensors have the same estimation distribution $\eta=[0.135, 0.239, 0.232, 0.394]$ and the same packet-loss probability, i.e.,  $\boldsymbol{\varepsilon}=[\varepsilon_{n,1},\varepsilon_{n,2},\varepsilon_{n,3},\varepsilon_{n,4}]$. Transmission in state $q$ requires $\omega_q=q$. According to \cite{igor_18_ton}, the optimum policy when all the sensors are identical without power constraint is to schedule the sensor with the largest AoI in each slot, which requires a minimum amount of $\mathcal{E}_G=\frac{M}{N}\sum_{q=1}^Q\eta_q\omega_q$ of power. We measure power constraint through factor $\rho_n=\mathcal{E}_n/\mathcal{E}_G$ and let $\rho_n=0.2+\frac{1.4}{N-1}n$. From the figure, the gap between the proposed algorithm and lower bound diminishes with $N$ increases.  

\begin{figure}[h]
	\centering
	\includegraphics[width=.4\textwidth]{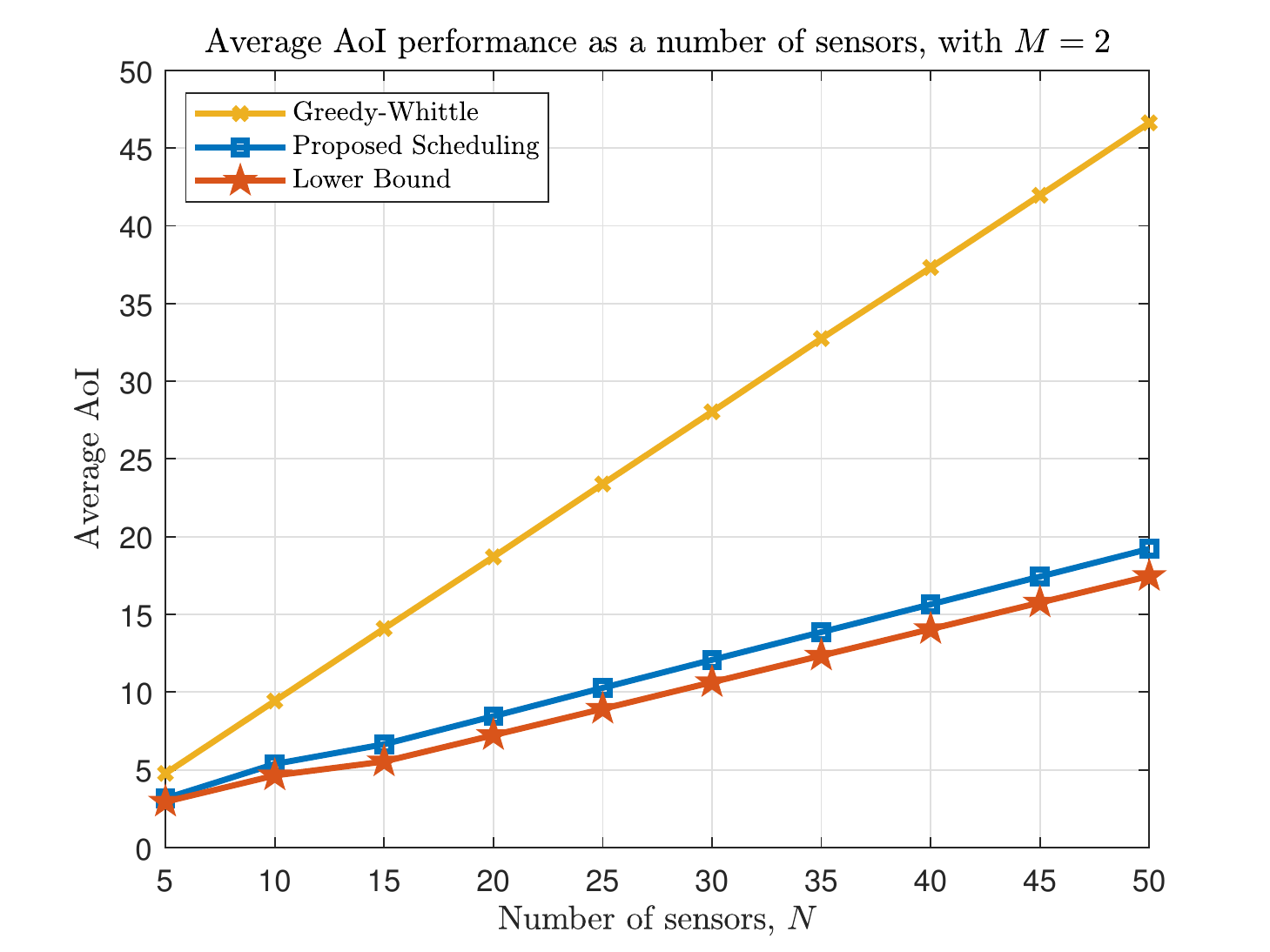}
	\caption{Average AoI performance in a network as a number of sensors. The bandwidth $M=2$ and packet-loss probability $\overline{\varepsilon}_n=\frac{n-1}{N}$. }
	\label{fig:AoI}
\end{figure}

Finally we compare the proposed algorithm with a naive Whittle's index approach. We consider a network where the packet-loss probability of each user is identical in every estimation, i.e., $\varepsilon_{n, q}=\overline{\varepsilon}_n$. With no power constraint, Whittle's index approach is shown to be near optimal to minimize AoI in bandwidth limited networks \cite{igor_18_ton}, and the index can be computed by:
\[W_n(x)=(1-\overline{\varepsilon}_n)x\left(x+\frac{1+\overline{\varepsilon}_n}{1-\overline{\varepsilon}_n}\right).\] At the beginning of each time slot, among all the sensors that have enough power, we choose no more than $M$ sensors with the largest Whittle's index and schedule them the send updates. This policy is denoted as Greedy-Whittle in Fig.~\ref{fig:AoI}. Our proposed scheduling policy achieves significant average AoI decrease compared with the greedy algorithm.
\section{Conclusion}
In this work, we study scheduling algorithm that minimizes average AoI performance in time-varying network when packet-loss may happen due to imperfect channel estimation. We solve the original scheduling problem through bandwidth relaxation and sensor decoupling. A truncated scheduling policy is proposed that can achieve asymptotic optimal performance while satisfies both bandwidth and power constraint. Our work suggests channel estimation imperfection can cause AoI growth. Sensors should be scheduled at higher AoI when the probability of packet-loss is large. 
\section*{Acknowledgment}
The authors are grateful to Dr.~Bo Zhou from Virginia Tech and Mr. Jingzhou Sun and Mr. Qining Zhang from Tsinghua University for fruitful discussions.
\appendix[Proof of Theorem 3]
Denote the policy that in each slot $t$, schedule sensor $n$ with probability $p_{x_n(t), q_n(t)}^{n, *}$ as policy $\pi_R$, which is the optimum policy that minimizes average AoI under relaxed bandwidth Eq.~\eqref{eq:relaxedbandwidth} and power constraint Eq.~\eqref{eq:powerconstraint}. Denote $\overline{\mathcal{I}}$ be the average number of sensors scheduled following policy $\pi_R$, since $\pi_R$ can satisfy the relaxed bandwidth constraint, we have:
\begin{equation}
	\overline{\mathcal{I}}=\mathbb{E}[\mathcal{I}(t)]\leq M.
\end{equation} Moreover, the average AoI following policy $\pi_R$ will be a lower bound of the average AoI of the primal scheduling problem, we have the following inequality:
\begin{equation}J(\hat{\pi})-J(\pi^*)\leq J(\hat{\pi})-J(\pi_R).\label{eq:proofeq1}
\end{equation}

We analyze the difference between $J(\hat{\pi})$ and $J(\pi_R)$ next. For sensor $n$ in slot $t$, the probability that it is scheduled following policy $\pi_R$ but is not scheduled following policy $\hat{\pi}$ is at most $z=\frac{N-M}{N}$. Notice that for AoI satisfies $x_n(t)>X$, it is always optimum to schedule the sensor. Then the probability that the sensor is still not scheduled in a consecutive $k$ slots is less than $z^{(k-(X-x_n(t)))^+}$. As a result, for $x_n(t)=x$, following policy $\hat{\pi}$ will cause an extra AoI of at most:
\begin{align}
a_x=&\sum_{k=0}^\infty(x_n(t)+k)z^{(k-(X-x))^+}\nonumber\\
\leq&\sum_{k=0}^{X-1}(x+k)+\sum_{k=X}^{\infty}(x+k)z^{k-X}\nonumber\\
\leq&xX+\frac{X(X-1)}{2}+\frac{1}{1-z}(x+X)+\frac{z}{(1-z)^2}
\end{align}

For simplicity, denote $C_1=\frac{X(X-1)}{2}+\frac{1}{1-z}X+\frac{z}{(1-z)^2}$ and $C_2=X+\frac{1}{1-z}$. Let us then consider another scheduling policy $\tilde{\pi}$, in each slot if policy $\pi_R$ schedules more than $M$ sensors, i.e., $|\mathcal{I}(t)|\geq M$, policy $\tilde{\pi}$ schedules all of them, but adds a penalty AoI of $a_{x_n(t)}$ on all the scheduled sensors. The average AoI following policy $\tilde{\pi}$ will be larger than following the truncated policy $\hat{\pi}$. Then, we have the following inequalities:
\begin{align}
	&J(\hat{\pi})-J(\pi_R)\nonumber\\
	\leq&J(\tilde{\pi})-J(\pi_R)\nonumber\\
	=&\mathbb{E}_{\pi_R}\left[\frac{1}{NT}\sum_{t=1}^T\sum_{n=1}^N a_{x_n(t)}(|\mathcal{I}(t)|-M)^+\right]\nonumber\\
	\overset{(a)}{\leq}&\mathbb{E}_{\pi_R}\left[\frac{1}{NT}\sum_{t=1}^T\sum_{n=1}^N a_{x_n(t)}(|\mathcal{I}(t)|-\overline{\mathcal{I}})^+\right]\nonumber\\
	\leq&\mathbb{E}_{\pi_R}\left[\frac{1}{NT}\sum_{t=1}^T\sum_{n=1}^N(C_2x_n(t)+C_1)||\mathcal{I}(t)|-\overline{\mathcal{I}}|\right]\label{eq:pf2},
\end{align}
where inequality (a) is due to Eq.~\eqref{eq:proofeq1}. Notice that for $x_n(t)>X$, the optimum policy is to schedule the sensor. Thus, the probability that the sensor's AoI larger than $X+k$ should be less than or equal to $(\max_{q}\varepsilon_{n, q})^k$. Denote $\rho=\max_{n, q}\varepsilon_{n, q}$, then for every $\epsilon$, there exists a $K=
\lceil\frac{\log\epsilon}{\log\rho}\rceil$, such that for any $x>X+K$, probability $\mu_x^{n, *}\leq\epsilon \rho^{x-(X+K)}$. Next, we proceeds to upper bound Eq.~\eqref{eq:pf2} by:
\begin{align}
	&\mathbb{E}_{\pi_R}\left[\frac{1}{NT}\sum_{t=1}^T\sum_{n=1}^N(C_2x_n(t)+C_1)||\mathcal{I}(t)|-\overline{\mathcal{I}}|\right]\nonumber\\
	=&\mathbb{E}_{\pi_R}\left[\frac{1}{NT}\sum_{t=1}^T\sum_{n=1}^N(C_1+C_2x_n(t)\mathbbm{1}_{x_n(t)\leq X+K})||\mathcal{I}(t)|-\overline{\mathcal{I}}|\right]\nonumber\\
	&+\mathbb{E}_{\pi_R}\left[\frac{1}{NT}\sum_{t=1}^T\sum_{n=1}^NC_2x_n(t)\mathbbm{1}_{x_n(t)> X+K}||\mathcal{I}(t)|-\overline{\mathcal{I}}|\right]\nonumber\\
	=&(C_1+C_2(X+K))\mathbb{E}_{\pi_R}\left[\sum_{t=1}^T\frac{1}{T}||\mathcal{I}(t)|-\overline{\mathcal{I}}|\right]\nonumber\\
	&+\mathbb{E}_{\pi_R}\left[\frac{1}{NT}\sum_{t=1}^T\sum_{n=1}^NC_2Nx_n(t)\mathbbm{1}_{x_n(t)>X+K}\right]\nonumber\\
	\leq&(C_1+C_2(X+K))\mathbb{E}_{\pi_R}\left[\sum_{t=1}^T\frac{1}{T}||\mathcal{I}(t)|-\overline{\mathcal{I}}|\right]\nonumber\\
	&+N\epsilon((X+K)\frac{1}{1-\rho}+\frac{\rho}{(1-\rho)^2})\label{eq:pf3}
\end{align}

According to \cite{diaconis1991closed}, $\mathbb{E}_{\pi_R}\left[\sum_{t=1}^T\frac{1}{T}||\mathcal{I}(t)|-\overline{\mathcal{I}}|\right]=\mathcal{O}(\frac{1}{\sqrt{N}})$. By choosing $\epsilon=N^{-2}$ and thus $K=\mathcal{O}(\log N)$ we can then upper bound Eq.~\eqref{eq:pf3} by:
\[\mathcal{O}(\frac{C_1+C_2(X+\log N))}{\sqrt{N}})+\mathcal{O}(\frac{\frac{1}{\varepsilon}(X+\log N)}{N})\]

Since $M/N$ keeps a constant, the scheduling threshold $X$ and coefficient $C_1, C_2$ do not increase with $N$. Thus the proposed scheduling strategy is asymptotic optimum. 
	\bibliography{bibfile}
\end{document}